\documentclass[sigconf,nonacm]{acmart}

\usepackage{todonotes}
\usepackage{subcaption}

\AtBeginDocument{%
  \providecommand\BibTeX{{%
    \normalfont B\kern-0.5em{\scshape i\kern-0.25em b}\kern-0.8em\TeX}}}

\copyrightyear{}
\acmYear{}
\acmDOI{}

\acmConference[]{}
\acmISBN{}

\begin{document}

\title{iNNk: A Multi-Player Game to Deceive a Neural Network}


\author{Jennifer Villareale$^{1}$, Ana Acosta-Ruiz$^{1}$, Samuel Arcaro$^{1}$, Thomas Fox$^{1}$, Evan Freed$^{1}$, Robert Gray$^{1}$, Mathias L{\"o}we$^{2}$, Panote Nuchprayoon$^{1}$, Aleksanteri Sladek$^{1}$, Rush Weigelt$^{1}$, Yifu Li$^{1}$, Sebastian Risi$^{2}$, Jichen Zhu$^{1}$}
\affiliation{%
  \institution{$^{1}$Drexel University, Philadelphia, USA\\
  $^{2}$IT University of Copenhagen, Copenhagen, Denmark}
  {\{jmv85, ava48, saa35, tbf33, emf67, rcg48, pn355, ams939, rw643, yl3374, jichen\}}@drexel.edu, {\{malw, sebr\}}@itu.dk
}
\renewcommand{\shortauthors}{Villareale, et al.}

\begin{abstract}
This paper presents \textit{iNNK}, a multiplayer drawing game where human players team up against an NN. The players need to successfully communicate a secret code word to each other through drawings, without being deciphered by the NN. With this game, we aim to foster a playful environment where players can, in a small way, go from passive consumers of NN applications to creative thinkers and critical challengers. 


\end{abstract}



\maketitle

\section{Introduction}
\textit{iNNK} is a multi-player game about deceiving artificial intelligence (AI). 
Recently AI technology, especially artificial neural networks (ANN or NN), has made significant leaps to recognize and generate images. At the 2015 ImageNet Large Scale Visual Recognition Challenge, AIs built by {\em Microsoft} and {\em Google} beat humans at image recognition tasks. Before the competition, NN's image recognition capability has already been used in everyday social situations. For instance, they were used to find suspicious behavior patterns during a user's computer usage \cite{stahl2010ethical} and be utilized in large public locations to survey individuals \cite{kamgar2011toward}. 

Artists and designers have experimented with ways to critically question the deployment of image recognition and to subvert its use through playful means. For instance,  Seoul-based designer Sang Mun designed {\em ZXX Typeface}\footnote{\url{https://walkerart.org/magazine/sang-mun-defiant-typeface-nsa-privacy}}, letters deliberately designed to be difficult to be recognized by computers, to protect users' privacy. In the project {\em Computer Vision Dazzle}\footnote{\url{https://cvdazzle.com/}}, artist Adam Harvey explores how fashion and makeup can be used as camouflage from face-detection technology. This project adopts a similarly playful approach and invites players to develop their own strategies to defy an NN in visual communication.

This paper presents \textit{iNNK}, a drawing game where two or more people play together with an NN. To win the game, the players need to successfully communicate a secret code word to each other through drawings, without being deciphered by the NN-based AI. Each game is composed of five short rounds, allowing players to experiment with different drawing strategies. This paper also summarizes the main strategies our players have developed in our playtesting. Certainly, a lot more effort is needed to empower citizens to be more familiar with AI and to engage the technology critically. Through our game, we have seen evidence that playful experience can turn people from passive users into creative and reflective thinkers, a crucial step towards a more mature relationship with AI.

\section{Related Work}
\label{rw}

With recent breakthroughs in neural networks (NN), particularly deep learning, designers are increasingly exploring their use in computer games. For example, researchers use NNs to procedurally generate game content, which otherwise would have to be created by human artists and thus make the game more variable ~\cite{risi2015petalz,yang2018learning}. 

NNs are also used to provide complex behavior for non-player characters (NPCs). For example, in the game \textit{Supreme Commander 2}~\cite{rabin2015game}, players combat against an army controlled by an NN. As the player customizes her army, the NN observes the player's unit composition and makes battle decisions, such as how its army will respond, which enemy to target first, or when to retreat. In this case, the NN makes gameplay more personalized and potentially more engaging.

Recently, Google's \textit{Quick, Draw!} \cite{quick}, an NN-based drawing detection demo, was developed to help with machine learning research. Users are given a keyword to draw, and their goal is to draw in a way that the NN can recognize it. Through the vast amount of data collected in the demo, Google retrains its NN and improves its accuracy. Our game was inspired by {\em Quick, Draw!}, and we added strong game mechanics to encourage rivalry between human players and the NN.




Many projects use an NN to provide playable experiences. For example, \textit{How to Train your Snake} \cite{snake}, modeled after the original game {\em Snake}, contains four independent snakes that can grow in length by eating food. Every snake is controlled by an NN. To win the game, the player must steer the NN progress through various upgrades to make the snakes reach a particular length. \textit{How to Train your Snake} \cite{snake} explicitly calls players' attention to the NN and turns the gameplay into a puzzle about how to best configure the NN. This design creates a setting for players to experiment with the NN. Through play, players may be more likely to understand the system's capabilities. With few exceptions (e.g., {\em Black \& White} \cite{wexler2002artificial}, \emph{Creatures} \cite{grand1997creatures}, \emph{Forza Car Racing} \cite{takahashi_2018}), \emph{GAR} \cite{hastings2009evolving}, {\em Hey Robot} \cite{heyrobo}, and \emph{Quick Draw!} \cite{quick}), most of these projects exist in the continuum between a tech demo and fully-fledged game. In \textit{iNNK}, we attempt to push the NN-based interactive experience to the latter.  

In related NN-based games, the player-NN interaction tends to be assistive --- either the players control the NN \cite{heyrobo,semantris,snake,wexler2002artificial,grand1997creatures} or the NN improves the game for the player \cite{hastings2009evolving,risi2015petalz,aidungeon}. Here, we create an adversarial relationship between the human players and the NN and thus encourage the players to be more creative competitors.

\section{Designing {\em iNNk}}
This section describes the design rationale of {\em iNNk}.
\subsection{Game Overview}
\label{overview}
\textit{iNNk} is a web-based multiplayer drawing game where two or more people play together against an NN. To win the game, the players need to successfully communicate a secret code word to each other through drawings, without being deciphered by the NN. Each game has five 30-second rounds; whoever (i.e., humans or the NN) has the most points at the end of the five rounds wins the match. For the human players to be successful, they must develop drawing strategies that can only be interpreted by their human teammates and not the NN.

\begin{figure*}[!t]
  \includegraphics[width=\textwidth]{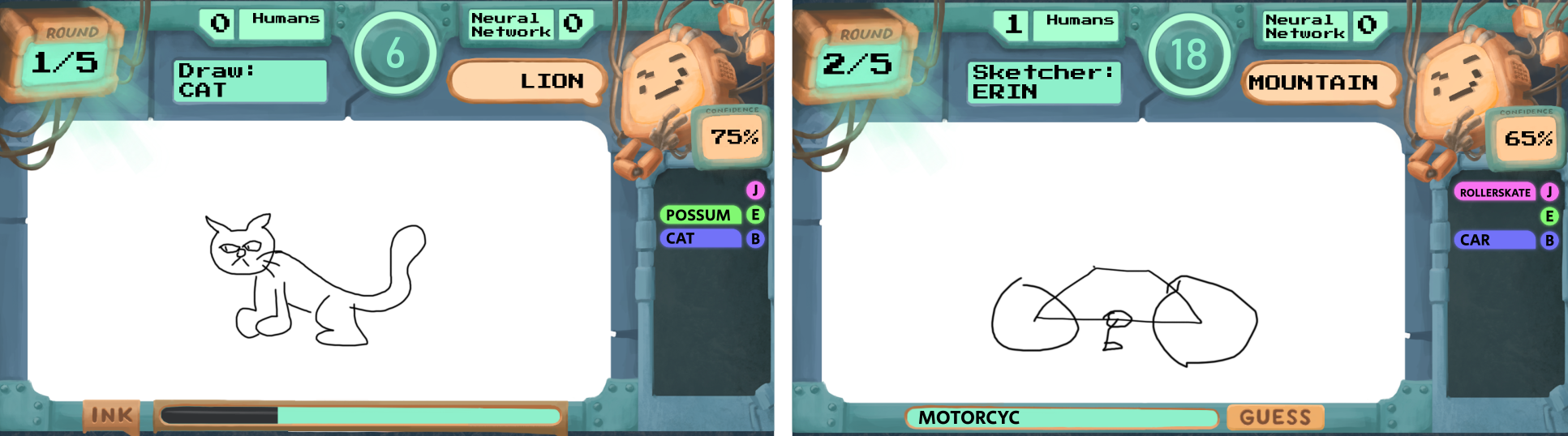}
  \caption{left: A screenshot of the Sketcher's interface. In the white canvas, the Sketcher draws to communicate the secret code word, indicated above the canvas ({\em cat}). The NN's guess and its confidence are on the upper right corner. Right: A screenshot of the Guesser's interface. Guessers can type in their guess at the bottom.} 
\label{fig:sketcher}
\end{figure*}

Players are assigned one of the two roles during the game: the Sketcher and the Guesser (Figure \ref{fig:sketcher}). The Sketcher is tasked with drawing something based on the code word assigned by the game. The goal is to draw the code word in such a way as the human Guesser may be able to accurately interpret the code word before the NN. In general, if the Sketcher draws something that is prototypical, it would be straightforward to other human players. But it will also be easy for the NN to guess. 

The Guessers are tasked with entering their guess of the code word based on the Sketcher's drawing before the NN guesses correctly. The NN always plays the role of one of the Guessers, and its goal is to decipher correctly first. While there is no penalty for wrong guesses, the human guessers must be mindful of their guesses. As described below in section \ref{nn}, the NN takes into account all previous attempts at the code word. Human Guessers' wrong guesses will increase the NN's chance for correct interpretation. 

The game is structured around five 30-second rounds. If either the Human team (consisting of one Sketcher and at least one Guesser) or the NN guesses the code word correctly within 30 seconds, the respective side wins the round. Otherwise, the round restarts the countdown until one side wins that round. Whichever team gets the most points at the end of round five wins the match. We chose to use 30-second rounds because it provides enough time for players to draw (or interpret) the code word while being quick enough to encourage frequent moments of surprise and failure. Ultimately, we intend for this to provoke explorations of different drawing strategies and encourage the players to think creatively about how to defy the AI.  

We also include an Ink Meter, which can be seen on the bottom of the left image in Figure \ref{fig:sketcher}. The purpose of the Ink Meter is to limit the amount a Sketcher can draw during the round. A common strategy observed from playtesting (see section \ref{expose} for more details) was the inclusion of visual noise (e.g., crosshatching) or distractions (e.g., other shapes) in addition to the drawing of the code word. We use the Ink Meter as a balancing tool to match the capabilities of the Sketcher and of the NN. This way, one side will not dominate the game easily and therefore create a more engaging player experience.  

\textit{iNNk's} intended audience is the general public, especially players who are interested in NNs. As a multiplayer game, we intend this to be played in a group setting, where players are encouraged to discuss between each round and collaboratively develop different strategies. 
We acknowledge that our design currently does not account for the possibility for players to ``cheat'' by directly telling each other the code word outside the game. We made this design decision partly following the convention of similar multiplayer games such as {\em Pictonary} \cite{pictionary}, {\em Cranium} \cite{cranium}, and {\em Charades} \cite{charades}. These games assume that players are more incentivized to have a good game than to win too easily. In addition, we want to encourage players to co-develop creative strategies to defeat the AI. We believe that leaving their communication channel open is a good way to accomplish this goal.

\subsection{How the Neural Network Works}
\label{nn}
In \textit{iNNk}, we use deep learning as the framework for our AI. Specifically, we reference Google's \textit{Quick Draw!} \cite{quick} architecture that leverages a sophisticated NN architecture with a combination of convolutional and LSTM layers, as well as batch normalization and dropout as regularization techniques. Our model was trained on hand-labeled sketch data from a canvas similar to the one used in our game. This data was taken from Google’s publicly available \textit{Quick Draw!} \cite{quick} dataset and includes 50 million drawings across 345 categories (i.e., 345 supported secret code words) of example sketches. 

Once trained, our model starts to make predictions (i.e., internal guesses) from the moment when the Sketcher makes the first stroke on the canvas. The categorization label with the highest predicted confidence constitutes the guess of the NN. The NN continues to generate guesses, however, they are only presented to the player once it is above a certain confidence value. Previous, incorrect guesses by both the NN and the human players are used to mask the output of the NN by removing these categories before rendering the guess of the NN. In this way, the NN is able to participate in a way that mimics the other human Guessers.

As discussed above, \textit{iNNk} incorporates an Ink Meter feature. We determine the depletion of ink as the Sketcher draws by defining a maximum length that drawn strokes can cumulatively cover across the canvas. As different drawings require different amounts of ink to be adequately represented (e.g., the "line" code word can be sufficiently represented with much less ink than the "car" code word), this maximum length is defined separately for each drawing. We reference the \textit{Quick, Draw!} \cite{quick} dataset's drawings to calculate an average distance that each category of drawings cover to determine Ink Meter values for \textit{iNNk's} drawings. 

\subsection{Highlighting the NN's Presence}

Many existing NN-based games obscure the existence of the NN. NNs are often seen as an underlying tool that the player only interacts with indirectly. For example, in the game \textit{Black \& White} \cite{wexler2002artificial}, players directly interact with the Creature, without the knowledge that it is controlled by an NN.


\begin{figure*}[!t]
  \includegraphics[width=\textwidth]{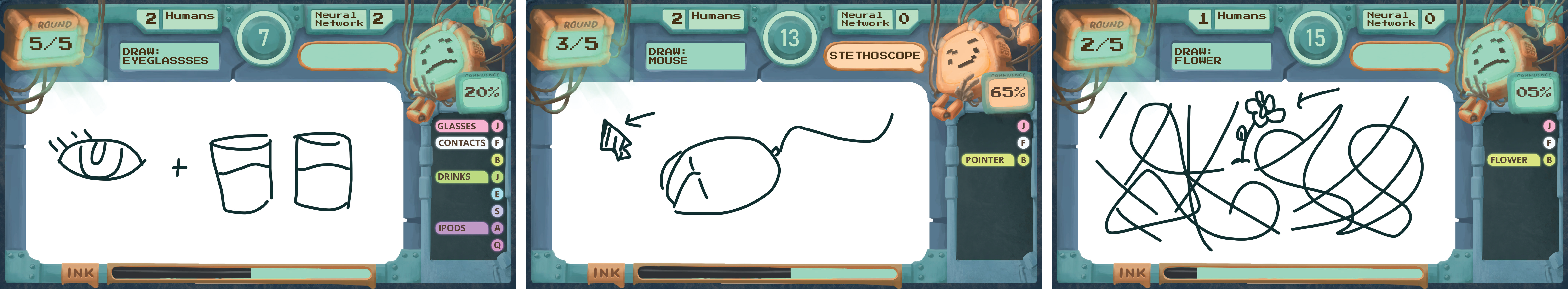}
  \caption{The image displays the three observed drawing strategies: sequential set of drawings, non-obvious drawing, and visual noise.} 
\label{fig:strategies}
\end{figure*}

Since our conceptual goal is to encourage people to go from passive users of NNs to active and creative challengers, we made the design decision to highlight the NN's presence as is. Rather than using a more abstract metaphor, we present the NN algorithm as a neural network non-player character (NPC) opponent. We specifically designed the NPC to look like a characterized computer system. We also explicitly call the character "Neural Network" in the core GUI and expose its confidence value to further emphasize the functional aspects of the system. The confidence value of the NN is displayed as a percentage under the NPC character, which can be seen in the top right image of Figure \ref{fig:sketcher}, to draw players' attention to how certain the NN is in their current guess. For example, when the NN becomes more certain, the character's screen color changes, and the confidence value increases. This provides players with a visual indication of when the NN may correctly guess the code word. From the Sketcher's perspective, knowing which line she just drew significantly increased the NN's confidence will help her build a better mental model of how the technology works and thus how to subvert it. 

\subsection{Player-NN Interaction}


In our game, players interact with the NN through an adversarial game setting where the players' opponent is the NN, which provides a consistent and playful reminder of the NN in the core gameplay loop. This setting allows players to consistently see the NN's output (i.e., guess and confidence value), reflect on it, and use this information to better gauge their drawing or guesses. As opposed to being in control of its output, as seen in \textit{How To Train Your Snake} \cite{snake}, players are able to link the NN's output to their current gameplay to better develop strategies to stump or trick the NN. 


Additionally, this setting creates a leveled power structure where both the players and the NN share the same resources, information, and game objective. All guesses are displayed in the UI for all human players and the NN to use to determine future guesses and drawing adjustments. Both the player and NN aim to interpret the drawing to determine the code word as quickly as possible. 

\subsection{Creating Moments of Surprise and Failure}
Many game designers use repeated failure to encourage players to rethink their gameplay strategies \cite{juul2013art}. Through failure, players can reflect on their current actions \cite{juul2013art}, self-correct, and use these experiences to become better in the game \cite{anderson2018failing,gee2003video}.

We attempt to create a playful experience through our design, where human players are encouraged to become more active and creative challengers against a NN. Through moments of surprise and failure, \textit{iNNk} intends to provoke explorations of different drawing strategies and encourage the players to think creatively about how to defy the AI. 

We support these moments through the game structure (i.e., timed rounds) and the NN's confidence meter (i.e., NN confidence value) to facilitate player reflection and new drawing strategies. A moment of surprise in our game might be to "wow" players with the strengths of the NN's image recognition. For example, in most cases, the NN is exceptionally good at guessing the code word from a limited or incomplete drawing. This strength is intended to surprise and defeat the human players to trigger player reflection on how they may draw the code word differently for the next round. 

The confidence meter provides a visual reference for players to use when performing a strategy.  For example, if a player tries to slow the NN down by drawing extra lines before they begin to draw the code word, they can see this strategy take effect on the NN's confidence meter, by the percentage decrease.  As a result, players are able to link their gameplay (i.e., drawing and guesses) to what decreases the confidence value.  By visualizing this value in the interface, we intend to provide players with another source of visual feedback, in addition to winning or losing the round, on what strategies may impact the NN's certainty.

\section{Observed Player Strategies}
\label{expose}

Based on our playtesting, we observed the following three strategies commonly developed by players collectively in our game. These strategies can be seen in Figure \ref{fig:strategies}. The first strategy includes the Sketcher drawing the code word in a sequential set of images (i.e., as a rebus). In this case, the Sketcher was given the code word "eyeglasses," they sketched two separate images, an eye and a pair of drinking glasses, in an attempt to stump the NN. As a result, the NN did not successfully guess the code word.

The second strategy includes the Sketcher drawing the code word in a non-obvious way. Some words can have multiple meanings that the NN may not understand. In this case, the Sketcher was given the code word "mouse." The player drew the word as a computer mouse, as opposed to a small creature with ears and a tail. As a result, the NN guessed incorrectly, and the humans won the round.

The third strategy includes adding visual noise or other shapes in addition to the drawing of the code word. In this case, the Sketcher crosshatched or drew other shapes as to mislead or slow the NN down, giving their human guesser a chance to interpret their drawing. This strategy was the most common from our playtesting, which inspired our Ink Meter feature as a way to balance the over-use of drawing in our game. The intention was to encourage players to be mindful of how much they are drawing, as this can mislead the NN and also the other human players.

\section{Conclusion}
In this paper, we present \textit{iNNK}, an NN-based drawing game about deceiving AI. With this game, we aim to foster a playful environment where players can go from passive consumers of NN applications to creative thinkers and critical challengers. While this game only does so in a very small way, we hope with more designs and art projects like ours, our society can gradually evolve a more healthy relationship with AI.  



\bibliographystyle{ACM-Reference-Format}
\bibliography{references}


\begin{thebibliography}{20}


\ifx \showCODEN    \undefined \def \showCODEN     #1{\unskip}     \fi
\ifx \showDOI      \undefined \def \showDOI       #1{#1}\fi
\ifx \showISBNx    \undefined \def \showISBNx     #1{\unskip}     \fi
\ifx \showISBNxiii \undefined \def \showISBNxiii  #1{\unskip}     \fi
\ifx \showISSN     \undefined \def \showISSN      #1{\unskip}     \fi
\ifx \showLCCN     \undefined \def \showLCCN      #1{\unskip}     \fi
\ifx \shownote     \undefined \def \shownote      #1{#1}          \fi
\ifx \showarticletitle \undefined \def \showarticletitle #1{#1}   \fi
\ifx \showURL      \undefined \def \showURL       {\relax}        \fi
\providecommand\bibfield[2]{#2}
\providecommand\bibinfo[2]{#2}
\providecommand\natexlab[1]{#1}
\providecommand\showeprint[2][]{arXiv:#2}

\bibitem[\protect\citeauthoryear{Anderson, Dalsen, Kumar, Berland, and
  Steinkuehler}{Anderson et~al\mbox{.}}{2018}]%
        {anderson2018failing}
\bibfield{author}{\bibinfo{person}{Craig~G Anderson}, \bibinfo{person}{Jen
  Dalsen}, \bibinfo{person}{Vishesh Kumar}, \bibinfo{person}{Matthew Berland},
  {and} \bibinfo{person}{Constance Steinkuehler}.}
  \bibinfo{year}{2018}\natexlab{}.
\newblock \showarticletitle{Failing up: How failure in a game environment
  promotes learning through discourse}.
\newblock \bibinfo{journal}{\emph{Thinking Skills and Creativity}}
  \bibinfo{volume}{30} (\bibinfo{year}{2018}), \bibinfo{pages}{135--144}.
\newblock


\bibitem[\protect\citeauthoryear{Bewelge}{Bewelge}{2017}]%
        {snake}
\bibfield{author}{\bibinfo{person}{Bewelge}.} \bibinfo{year}{2017}\natexlab{}.
\newblock \bibinfo{title}{How to Train Your Snake}.
\newblock
\newblock
\urldef\tempurl%
\url{https://bewelge.itch.io/how-to-train-your-snake}
\showURL{%
\tempurl}


\bibitem[\protect\citeauthoryear{Gee}{Gee}{2003}]%
        {gee2003video}
\bibfield{author}{\bibinfo{person}{James~Paul Gee}.}
  \bibinfo{year}{2003}\natexlab{}.
\newblock \showarticletitle{What video games have to teach us about learning
  and literacy}.
\newblock \bibinfo{journal}{\emph{Computers in Entertainment (CIE)}}
  \bibinfo{volume}{1}, \bibinfo{number}{1} (\bibinfo{year}{2003}),
  \bibinfo{pages}{20--20}.
\newblock


\bibitem[\protect\citeauthoryear{Google}{Google}{2016}]%
        {quick}
\bibfield{author}{\bibinfo{person}{Google}.} \bibinfo{year}{2016}\natexlab{}.
\newblock \bibinfo{title}{Quick, Draw!}
\newblock
\newblock
\urldef\tempurl%
\url{https://quickdraw.withgoogle.com/}
\showURL{%
\tempurl}


\bibitem[\protect\citeauthoryear{Google}{Google}{2018}]%
        {semantris}
\bibfield{author}{\bibinfo{person}{Google}.} \bibinfo{year}{2018}\natexlab{}.
\newblock \bibinfo{title}{Semantris}.
\newblock
\newblock
\urldef\tempurl%
\url{https://research.google.com/semantris/}
\showURL{%
\tempurl}


\bibitem[\protect\citeauthoryear{Grand, Cliff, and Malhotra}{Grand
  et~al\mbox{.}}{1997}]%
        {grand1997creatures}
\bibfield{author}{\bibinfo{person}{Stephen Grand}, \bibinfo{person}{Dave
  Cliff}, {and} \bibinfo{person}{Anil Malhotra}.}
  \bibinfo{year}{1997}\natexlab{}.
\newblock \showarticletitle{Creatures: Artificial life autonomous software
  agents for home entertainment}. In \bibinfo{booktitle}{\emph{Proceedings of
  the first international conference on Autonomous agents}}.
  \bibinfo{pages}{22--29}.
\newblock


\bibitem[\protect\citeauthoryear{Hastings, Guha, and Stanley}{Hastings
  et~al\mbox{.}}{2009}]%
        {hastings2009evolving}
\bibfield{author}{\bibinfo{person}{Erin~J Hastings}, \bibinfo{person}{Ratan~K
  Guha}, {and} \bibinfo{person}{Kenneth~O Stanley}.}
  \bibinfo{year}{2009}\natexlab{}.
\newblock \showarticletitle{Evolving content in the galactic arms race video
  game}. In \bibinfo{booktitle}{\emph{2009 IEEE Symposium on Computational
  Intelligence and Games}}. IEEE, \bibinfo{pages}{241--248}.
\newblock


\bibitem[\protect\citeauthoryear{Juul}{Juul}{2013}]%
        {juul2013art}
\bibfield{author}{\bibinfo{person}{Jesper Juul}.}
  \bibinfo{year}{2013}\natexlab{}.
\newblock \bibinfo{booktitle}{\emph{The art of failure: An essay on the pain of
  playing video games}}.
\newblock \bibinfo{publisher}{MIT press}.
\newblock


\bibitem[\protect\citeauthoryear{Kamgar-Parsi, Lawson, and
  Kamgar-Parsi}{Kamgar-Parsi et~al\mbox{.}}{2011}]%
        {kamgar2011toward}
\bibfield{author}{\bibinfo{person}{Behrooz Kamgar-Parsi},
  \bibinfo{person}{Wallace Lawson}, {and} \bibinfo{person}{Behzad
  Kamgar-Parsi}.} \bibinfo{year}{2011}\natexlab{}.
\newblock \showarticletitle{Toward development of a face recognition system for
  watchlist surveillance}.
\newblock \bibinfo{journal}{\emph{IEEE Transactions on Pattern Analysis and
  Machine Intelligence}} \bibinfo{volume}{33}, \bibinfo{number}{10}
  (\bibinfo{year}{2011}), \bibinfo{pages}{1925--1937}.
\newblock


\bibitem[\protect\citeauthoryear{Lantz}{Lantz}{2019}]%
        {heyrobo}
\bibfield{author}{\bibinfo{person}{Frank Lantz}.}
  \bibinfo{year}{2019}\natexlab{}.
\newblock \bibinfo{title}{Hey Robot}.
\newblock
\newblock
\urldef\tempurl%
\url{https://www.kickstarter.com/projects/924858949/hey-robot}
\showURL{%
\tempurl}


\bibitem[\protect\citeauthoryear{Mindware}{Mindware}{1990}]%
        {charades}
\bibfield{author}{\bibinfo{person}{Mindware}.} \bibinfo{year}{1990}\natexlab{}.
\newblock \bibinfo{title}{Family Charades In-A-Box}.
\newblock \bibinfo{howpublished}{Board Game}.
\newblock


\bibitem[\protect\citeauthoryear{Rabin}{Rabin}{2015}]%
        {rabin2015game}
\bibfield{author}{\bibinfo{person}{Steven Rabin}.}
  \bibinfo{year}{2015}\natexlab{}.
\newblock \bibinfo{booktitle}{\emph{Game AI pro 2: collected wisdom of game AI
  professionals}}.
\newblock \bibinfo{publisher}{AK Peters/CRC Press}.
\newblock


\bibitem[\protect\citeauthoryear{Risi, Lehman, D'Ambrosio, Hall, and
  Stanley}{Risi et~al\mbox{.}}{2015}]%
        {risi2015petalz}
\bibfield{author}{\bibinfo{person}{Sebastian Risi}, \bibinfo{person}{Joel
  Lehman}, \bibinfo{person}{David~B D'Ambrosio}, \bibinfo{person}{Ryan Hall},
  {and} \bibinfo{person}{Kenneth~O Stanley}.} \bibinfo{year}{2015}\natexlab{}.
\newblock \showarticletitle{Petalz: Search-based procedural content generation
  for the casual gamer}.
\newblock \bibinfo{journal}{\emph{IEEE Transactions on Computational
  Intelligence and AI in Games}} \bibinfo{volume}{8}, \bibinfo{number}{3}
  (\bibinfo{year}{2015}), \bibinfo{pages}{244--255}.
\newblock


\bibitem[\protect\citeauthoryear{Robert~Angel}{Robert~Angel}{1994}]%
        {pictionary}
\bibfield{author}{\bibinfo{person}{Gary~Everson Robert~Angel}.}
  \bibinfo{year}{1994}\natexlab{}.
\newblock \bibinfo{title}{Pictonary}.
\newblock \bibinfo{howpublished}{Board Game}.
\newblock


\bibitem[\protect\citeauthoryear{Stahl, Elizondo, Carroll-Mayer, Zheng, and
  Wakunuma}{Stahl et~al\mbox{.}}{2010}]%
        {stahl2010ethical}
\bibfield{author}{\bibinfo{person}{Bernd Stahl}, \bibinfo{person}{David
  Elizondo}, \bibinfo{person}{Moira Carroll-Mayer}, \bibinfo{person}{Yingqin
  Zheng}, {and} \bibinfo{person}{Kutoma Wakunuma}.}
  \bibinfo{year}{2010}\natexlab{}.
\newblock \showarticletitle{Ethical and legal issues of the use of
  computational intelligence techniques in computer security and computer
  forensics}. In \bibinfo{booktitle}{\emph{The 2010 International Joint
  Conference on Neural Networks (IJCNN)}}. IEEE, \bibinfo{pages}{1--8}.
\newblock


\bibitem[\protect\citeauthoryear{Takahashi}{Takahashi}{2018}]%
        {takahashi_2018}
\bibfield{author}{\bibinfo{person}{Dean Takahashi}.}
  \bibinfo{year}{2018}\natexlab{}.
\newblock \bibinfo{title}{How Microsoft's Turn 10 fashioned the A.I. for cars
  in Forza Motorsport 5 (interview)}.
\newblock
\newblock
\urldef\tempurl%
\url{https://venturebeat.com/2013/11/06/how-microsofts-turn-10-fashioned-the-ai-for-cars-in-forza-motorsport-5-interview/}
\showURL{%
\tempurl}


\bibitem[\protect\citeauthoryear{Walton}{Walton}{2019}]%
        {aidungeon}
\bibfield{author}{\bibinfo{person}{Nick Walton}.}
  \bibinfo{year}{2019}\natexlab{}.
\newblock \bibinfo{title}{AI Dungeon}.
\newblock
\newblock
\urldef\tempurl%
\url{https://aidungeon.io/}
\showURL{%
\tempurl}


\bibitem[\protect\citeauthoryear{Wexler}{Wexler}{2002}]%
        {wexler2002artificial}
\bibfield{author}{\bibinfo{person}{James Wexler}.}
  \bibinfo{year}{2002}\natexlab{}.
\newblock \showarticletitle{Artificial Intelligence in Games}.
\newblock \bibinfo{journal}{\emph{Rochester: University of Rochester}}
  (\bibinfo{year}{2002}).
\newblock


\bibitem[\protect\citeauthoryear{Whit~Alexander}{Whit~Alexander}{1997}]%
        {cranium}
\bibfield{author}{\bibinfo{person}{Richard~Tait Whit~Alexander}.}
  \bibinfo{year}{1997}\natexlab{}.
\newblock \bibinfo{title}{Cranium}.
\newblock \bibinfo{howpublished}{Board Game}.
\newblock


\bibitem[\protect\citeauthoryear{Yang and Onta{\~n}{\'o}n}{Yang and
  Onta{\~n}{\'o}n}{2018}]%
        {yang2018learning}
\bibfield{author}{\bibinfo{person}{Zuozhi Yang} {and} \bibinfo{person}{Santiago
  Onta{\~n}{\'o}n}.} \bibinfo{year}{2018}\natexlab{}.
\newblock \showarticletitle{Learning Map-Independent Evaluation Functions for
  Real-Time Strategy Games}. In \bibinfo{booktitle}{\emph{2018 IEEE Conference
  on Computational Intelligence and Games (CIG)}}. IEEE, \bibinfo{pages}{1--7}.
\newblock


\end{thebibliography}


\end{document}